# Enhancement of Noisy Speech Exploiting an Exponential Model Based Threshold and a Custom Thresholding Function in Perceptual Wavelet Packet Domain

Md Tauhidul Islam, Celia Shahnaz*, *Member, IEEE,* Wei-Ping Zhu, *Senior Member, IEEE,* and M. Omair Ahmad, *Fellow, IEEE*

## Abstract

For enhancement of noisy speech, a method of threshold determination based on modeling of Teager energy (TE) operated perceptual wavelet packet (PWP) coefficients of the noisy speech by exponential distribution is presented. A custom thresholding function based on the combination of $\mu$-law and semisoft thresholding functions is designed and exploited to apply the statistically derived threshold upon the PWP coefficients. The effectiveness of the proposed method is evaluated for car and multi-talker babble noise corrupted speech signals through performing extensive simulations using the NOIZEUS database. The proposed method outperforms some of the state-of-the-art speech enhancement methods both at high and low levels of SNRs in terms of the standard objective measures and the subjective evaluations including formal listening tests.

## I. INTRODUCTION

Determination of a signal that is corrupted by additive or multiplicative noise has been of interest because of its importance in both theoretical and practical field. The main interest is to recover the real signal from the noise-mixed data received from microphone, ecg machine, radar, mobile phone or any other sound devices. Our aim is to make the recreated signal close to the original one. The use of such operation has application in broad area of speech communication applications, such as mobile telephony, speech coding and recognition, and hearing aid devices [1], [2], [3].

Over the decades, several methods have been developed to solve the noise reduction and speech enhancement problem. We can divide these methods in mainly three categories based on their domains of operation, namely time domain, frequency domain and time-frequency domain. Time domain methods include the subspace approach [4], [5], frequency domain methods include methods based on discrete cosine transform [6], spectral subtraction [7], [8], minimum mean square error (MMSE) estimator [9], [10], wiener filtering [11], [12] and time frequency-domain methods involve the employment of the family of wavelets [13], [14], [15], [16], [17]. All these methods have their advantages and disadvantages. Time domain methods like subspace method provide a tradeoff between speech distortion and residual noise. But they need a large amount of computation and as a result real-time processing becomes very difficult with these methods. On the other hand, frequency domain methods provide the advantage of real-time processing with less computation load. Among frequency domain methods, the most prominent one is spectral subtraction method. This method provides the facility of deducting noise from the noisy signal based on stationary nature of noise in speech signals. But this method has a major drawback of producing an artifact named musical noise, which is perceptually disturbing, made of different tones of random frequencies and has an increasing variance. In the MMSE estimator, the spectral amplitude of noisy signal is modified based on the minimum square error. A large variance as well as worst performance in highly noisy situation are the main problems with this method. The main problem of wiener filter based methods is the necessity of clean speech statistics for their implementation. Like MMSE estimators, wiener filters also try to reach at an optimum solution depending on the error value between the computed signal with the real one.

The methods which use thresholding as process of removing noise, Universal thresholding [13], SURE [18], WPF [14] and BayesShrink [19] are the prominent ones. In Universal thresholding method, a common threshold derived from noise power is used to threshold the wavelet coefficients. SURE applies Steins Uncertainty and BayesShrink applies Bayes principle to determine the threshold. WPF is the modified version of Universal threshold method with speech and silent frame detection ability.

In this paper, modeling of the TE operated PWP coefficients of noisy speech by the exponential distribution is proposed in order to derive a threshold for performing the thresholding operation. Here, instead of direct employment of the TE operator on the noisy speech, the TE operator is applied on the PWP coefficients of the noisy speech for carrying out the statistical modeling and to determine the threshold. The threshold thus derived is adapted with respect to speech and silent frames. Finally, a thresholding function developed from the $\mu$-law and semisoft thresholding functions is employed on the PWP coefficients to obtain an enhanced speech.

The paper is organized as follows. Section II presents the Proposed Method. Section III describes results. Concluding remarks are presented in Section IV.

## II. PROPOSED METHOD

The block diagram for the proposed method is shown in Fig.1. It is seen from Fig.1 that PWP transform is first applied to each input speech frame. Then, the PWP coefficients are subject to TE operation with a view to determine a threshold value for performing thresholding operation in the PWP domain. On using a custom thresholding function, an enhanced speech frame is obtained via inverse perceptual wavelet packet (IPWP) transform.

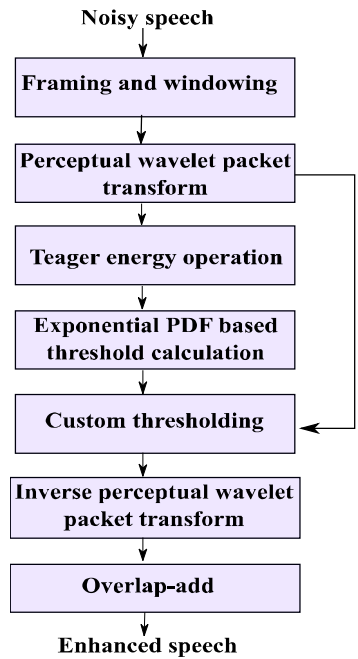

Fig. 1: Block diagram for the proposed method

### A. Perceptual Wavelet Packet Transform

The key element of the PWP transform is the use of the mel warping function to determine the wavelet packet decomposition structure [20]. The main motivation behind using this transform is its ability to decompose the signal according to human auditory system. At lower frequencies, where human auditory system can differentiate the pitches precisely, PWP transform decomposes the signal in finer bands. On the other hand, at higher frequencies, PWP transform creates less number of bands as the human cochlea can not differentiate small differences in higher frequencies. The perceptual mel scale is a scale of pitches judged by the listeners to be equal in distance from one another. The conversion of frequency to mel is discussed in [20].

The clean and noise PWP coefficients in a subband of a noisy speech frame at an SNR of 5dB is plotted in Fig.2. It is seen from this figure that for most of the coefficient indices, clean and noise PWP coefficients are not separable. Based on similar analysis performed on many speech signals corrupted by different noises, it is found that the time and frequency resolution provided by PWP transform is not sufficient to separate PWP coefficients of clean speech from that of noise even at a high SNR of 5dB. Since, TE operator has better time and frequency resolution [21], it can be very useful in handling noise. Therefore, we apply discrete time TE operator on the PWP coefficients.

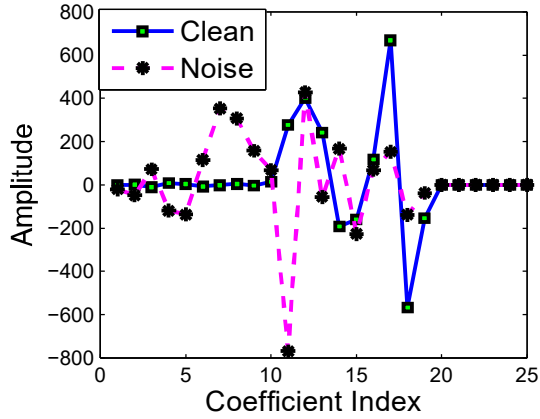

Fig. 2: PWP coefficients of a noisy speech subband at an SNR of 5dB

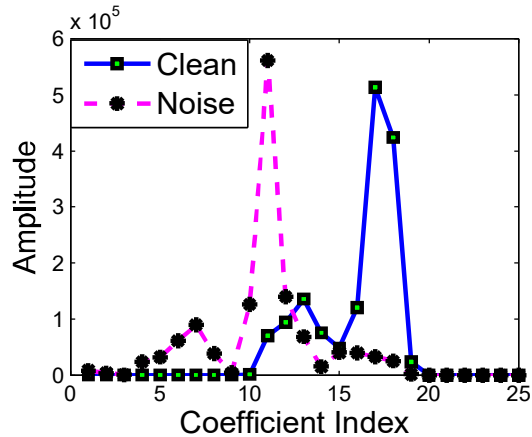

Fig. 3: TE operated PWP coefficients of a noisy speech subband at an SNR of 5dB

### B. Teager Energy Operator

Letting $W_{k,m}$ as the $m^{th}$ PWP coefficient in the $k^{th}$ subband, the TE operated coefficient $t_{k,m}$ corresponding to $W_{k,m}$ can be expressed as

$$t_{k,m} = T(W_{k,m}),  \tag{1}$$

where discrete TE operator $T(W_{k,m})$ is defined as [21]

$$T(W_{k,m}) = W_{k,m}^2 - W_{k,m+1}W_{k,m-1}.  \tag{2}$$

Fig. 3 presents the clean and noise TE operated PWP coefficients in a subband of a noisy speech frame at the same SNR as used in Fig. 2. It is seen from this figure that at the indices where TE operated PWP coefficients of clean speech have higher values, the TE operated PWP coefficients of noise show lower values. As a result, thresholding operation on the noisy PWP coefficients needs a low threshold value thus removing the noise leaving the speech undistorted. On the contrary, at the indices, where TE operated PWP coefficients of clean speech have lower values, the TE operated PWP coefficients of noise exhibit higher values as expected. Thus thresholding the noisy speech PWP coefficients needs a higher threshold value and removes the necessary noise without speech distortion at a significant level. Therefore, TE operation on PWP coefficients is found as more capable of serving the goal of thresholding operation by reducing the noise as well as preserving the speech.

### C. Proposed Model for TE Operated PWP Coefficients

The outcome of a speech enhancement method based on the thresholding in a transform domain depends mainly on two factors, namely the threshold value and the thresholding function. The use of a unique threshold for all the PWP subbands is not reasonable. As a crucial parameter, the threshold value in each subband is required to be adjusted very precisely so that it can prevent distortion in the enhanced speech as well as decrease annoying residual noise. By considering the probability distributions of $t_k$ of the noisy speech, noise and clean speech, a more accurate threshold value can be obtained using a suitable pattern matching scheme or similarity measure, where $t_k = t_{k,1}, \ldots, t_{k,M}$, $M$ is the total number of PWP coefficients in $k^{th}$ subband. Since speech is a time-varying signal, it is difficult to realize the actual probability distribution function (PDF) of speech or its $t_k$. As an alternative to formulate a PDF of the of speech, we can easily formulate the histogram of its $t_k$ and can approximate the histogram by a reasonably close PDF namely Gaussian and exponential. For TE operated PWP coefficients in a randomly chosen subband of a noisy speech frame, the empirical histogram along with the Gaussian and the exponential PDFs are superimposed in Fig. 4, 5 and 6 in presence of car noise at SNRs of $-15$, $0$ and $15$ dB. From these figures, it is obvious that exponential PDF fits the empirical histogram better than the Gaussian PDF. Similar analysis results are obtained for empirical histogram, Gaussian and exponential PDFs of TE operated noise PWP coefficients at a randomly chosen subband in speech signals of the same SNRs as used in Figs. 4, 5 and 6 and are shown in Figs. 7, 8 and 9. Such statistical matching of empirical histogram with the Gaussian and exponential PDFs is also explained in terms of AIC index [22]. It can be noted from [22] that the more negative value of the AIC index indicates the more matching between two PDFs. Assuming Gaussian and exponential PDFs for TE operated noise PWP coefficients in 100 randomly chosen subbands of a noisy speech frame, mean values of AIC indices obtained using different speech sentences are shown in Fig.10 for a range of SNR $-15$dB to $15$dB in the presence of car noise. From Fig.10, it is clearly attested that the exponential PDF offers better matching with the empirical histogram compared to the Gaussian PDF not only at SNR of 15dB but also at an SNR as low as $-15$dB. The plot representing the values of AIC indices for the Gaussian and exponential PDFs of TE operated noise PWP coefficients of noise in 100 randomly chosen subbands of speech files of SNR level ranging from $-15$dB to $15$dB is illustrated in Fig.11. This figure shows that AIC indices for $t_k$ of noise continues to exhibit smaller values for exponential PDF thus maintaining better matching with the empirical histogram for a wide range of SNR. Therefore, we propose to approximate the histograms of TE operated PWP coefficients of noisy speech, noise and clean speech by exponential PDF and use this approximation to calculate the threshold adaptive to different subbands.

### D. Determination of Proposed Adaptive Threshold

The entropy of each subband of the PWP coefficients is found different from each other. So, an entropy measure may be chosen to select a suitable threshold value adaptive to each subband. Some popular similarity measures those are related to the entropy functions are the Variational distance, the Bhattacharyya distance, the Harmonic mean, the Kullback Leibler(K-L) divergence, and the Symmetric K-L divergence. The K-L divergence is always nonnegative and zero if and only if the approximate exponential PDF of noisy speech and that of the noise or the approximate exponential PDF of the noisy speech and that of the clean speech are exactly the same. In order to have a symmetric distance between any two approximate exponential PDFs as mentioned above, the Symmetric K-L divergence has been adopted in this paper. The Symmetric K-L divergence is defined as

$$SKL(p,q) = \frac{KL(p,q) + KL(q,p)}{2},$$ (3)

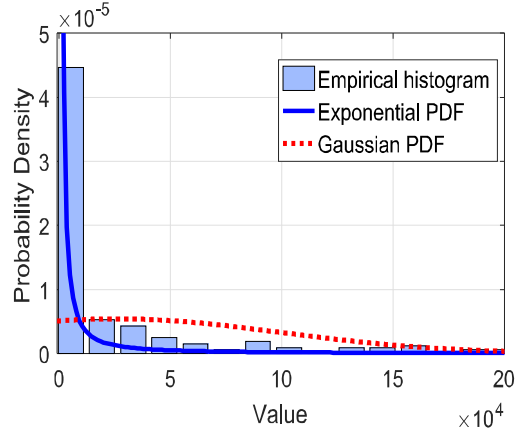

Fig. 4: Empirical histogram, Gaussian and exponential PDFs of TE operated PWP coefficients of noisy speech at SNR of −15 dB

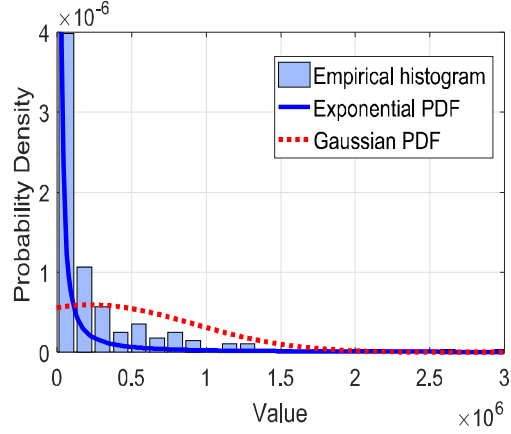

Fig. 5: Empirical histogram, Gaussian and exponential PDFs of TE operated PWP coefficients of noisy speech at SNR of 0 dB

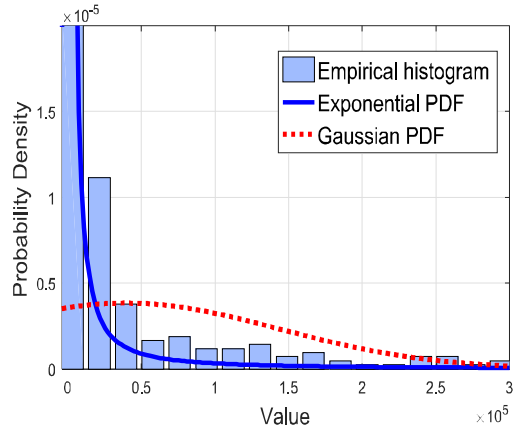

Fig. 6: Empirical histogram, Gaussian and exponential PDFs of TE operated PWP coefficients of noisy speech at SNR of 15 dB

where $p$ and $q$ are the two approximate exponential PDFs calculated from the corresponding histograms each having $N$ number of bins and $KL(p, q)$ is the K-L divergence between $p$ and $q$ given by

$$KL(p, q) = \sum_{i=1}^{N} p_i(t_k) \ln \frac{p_i(t_k)}{q_i(t_k)}. \tag{4}$$

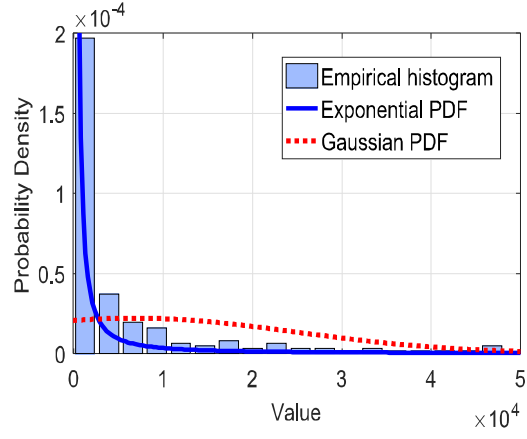

Fig. 7: Empirical histogram, Gaussian and exponential PDFs of TE operated noise PWP coefficients at SNR of $-15$ dB

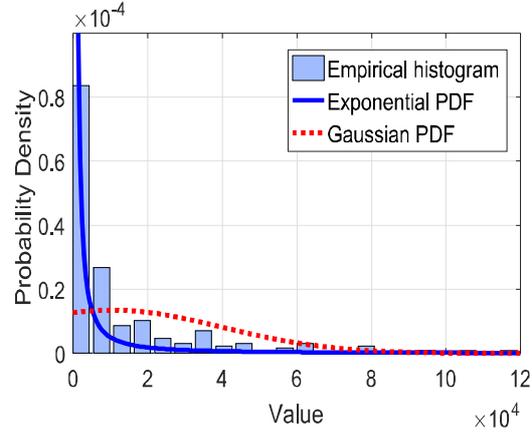

Fig. 8: Empirical histogram, Gaussian and exponential PDFs of TE operated noise PWP coefficients at SNR of $0$ dB

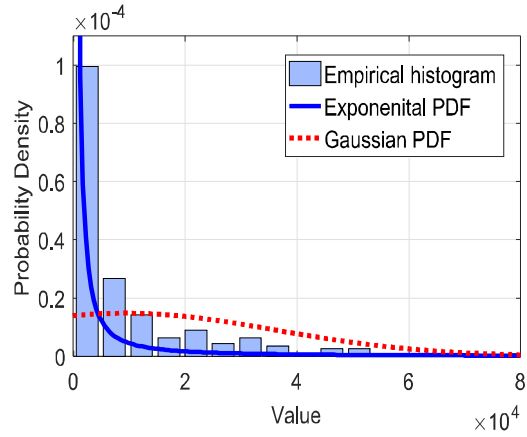

Fig. 9: Empirical histogram, Gaussian and exponential PDFs of TE operated noise PWP coefficients at SNR of $15$ dB

In (4), $p_i(t_k)$ is the probability of $t_k$ of noisy speech in $i^{th}$ bin given by

$$p_i(t_k) = \frac{n_i}{M},$$ (5)

where $n_i$ is number of coefficients in $i^{th}$ bin and $M$ total number of coefficients in $k^{th}$ subband. In this way, we can find the PDF for $t_k$ of noisy speech, $p_{t_k}$ and approximate the obtained PDF with the exponential PDF. We will denote the approximated exponential PDF with $\hat{p}_{t_k}$. Similarly, the approximate exponential PDF of the $t_k$ of the noise and that of the $t_k$ of the clean speech can be estimated following (5) and denoted by $\hat{q}_{t_k}$ and $\hat{r}_{t_k}$, respectively. Below a certain value of threshold $\lambda$, the symmetric K-L divergence between $\hat{p}_{t_k}$ and $\hat{q}_{t_k}$ is approximately zero, i.e.,

$$SKL(\hat{p}_{t_k}, \hat{q}_{t_k}) \approx 0. \tag{6}$$

Exponential PDF for $t_k$ of noisy speech can be written as

$$\hat{p}_{t_k}(x) = \frac{1}{\sqrt{\sigma_s^2}} \times \exp(-\frac{x}{\sqrt{\sigma_s^2}}), x \in [0, \infty], \tag{7}$$

where $\sigma_s^2$ represents the power of $t_k$ of noisy speech. Letting $\sigma_r^2$ as the power of $t_k$ of clean speech and $\sigma_n^2$ as the power of $t_k$ of noise and using the fact $\sigma_s^2 = \sigma_r^2 + \sigma_n^2$, we can write

$$\hat{p}_{t_k}(x) = \frac{1}{\sqrt{(\sigma_r^2 + \sigma_n^2)}} \times \exp(-\frac{x}{(\sqrt{\sigma_r^2 + \sigma_n^2})}), x \in [0, \infty]. \tag{8}$$

Following (7) in a similar way, exponential PDF for $t_k$ of noise can also be written as

$$\hat{q}_{t_k}(x) = \frac{1}{\sqrt{\sigma_n^2}} \times \exp(-\frac{x}{\sqrt{\sigma_n^2}}), x \in [0, \infty]. \tag{9}$$

By substituting (8) and (9) in (6), we obtain

$$\int_0^\lambda [\frac{1}{\sqrt{\sigma_s^2}} \times \exp(-\frac{x}{\sqrt{\sigma_s^2}}) - \frac{1}{\sqrt{\sigma_n^2}} \times \exp(-\frac{x}{\sqrt{\sigma_n^2}})] I_1 dx = 0, \tag{10}$$

where $I_1$ is defined as $I_1 = \ln \sqrt{\frac{\sigma_n^2}{\sigma_s^2}} \times \exp(-\frac{x}{\sigma_s^2} + \frac{x}{\sigma_n^2})$. By solving (10), value of $\lambda(k)$ can be obtained as

$$\lambda(k) = \frac{\sqrt{\sigma_n^2(k)(1 + \gamma(k))} \times \ln(\sqrt{1 + \gamma(k)})}{\sqrt{1 + \gamma(k)} - 1}, \tag{11}$$

where $\gamma(k)$ is the SNR of subband $k$ defined as

$$\gamma(k) = \frac{\sigma_r^2(k)}{\sigma_n^2(k)}. \tag{12}$$

The proposed threshold $\lambda(k)$ in (11) derived assuming exponential PDF is compared with that obtained assuming Gaussian PDF given by [24]

$$\lambda(k) = \sqrt{\sigma_n^2(k)}\sqrt{2(\gamma_k + \gamma_k^2)} \times \ln(\sqrt{1 + \frac{1}{\gamma_k}}) \tag{13}$$

in Fig.12. This figure shows that the pattern of the threshold value is similar for both the PDFs. In terms of value, although exponential PDF shows slightly higher values at higher SNRs, but the threshold values are much lower than that of the Gaussian PDF specially at lower SNRs. Therefore, the threshold derived from the exponential PDF not only offers less chance of removing speech coefficients while performing thresholding operation at lower SNRs, but also ensures better removal of noise at higher SNRs.

The proposed threshold as derived in (11) is high for higher noise power and low for lower noise power thus is adaptive to noise power of different subbands. In this method, voice activity detector is not necessary as the threshold is automatically adapted to the silent and speech frames. At a silent frame, since noise power is significantly higher than the signal power, the proposed threshold results in a higher value as seen from (11). Such a value imposes more coefficients to be thresholded to zero thus removes noise coefficients completely at subbands of a silent frame. Note that, in this paper, noise is estimated using Improved Minima Controlled Recursive Averaging (IMCRA) method [23].

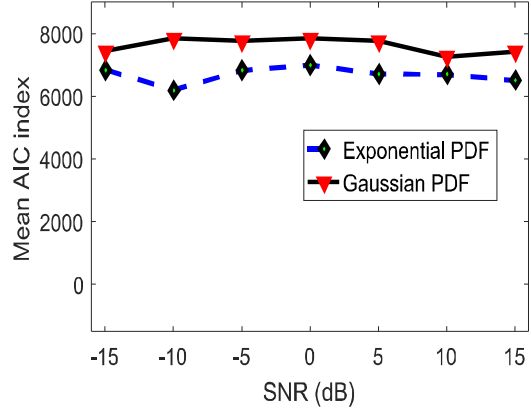

Fig. 10: Mean values of AIC indices of TE operated PWP coefficients of noisy speech assuming Gaussian and exponential PDFs

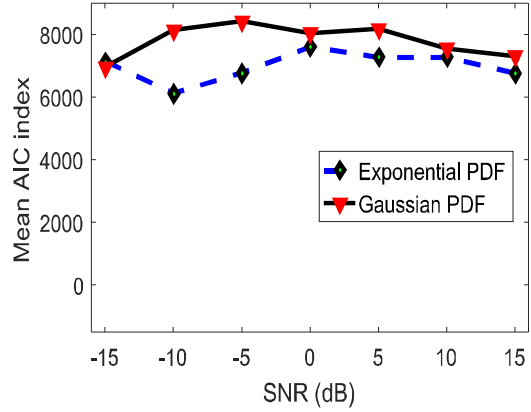

Fig. 11: Mean values of AIC indices of TE operated PWP coefficients of noise assuming Gaussian and exponential PDFs

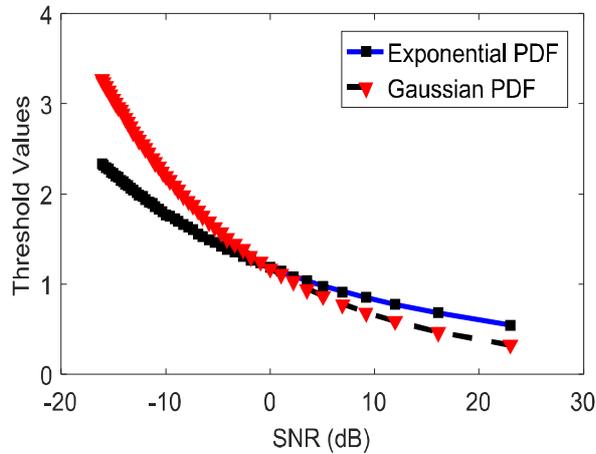

Fig. 12: Comparison of threshold values with respect to SNR for exponential and Gaussian PDFs

### E. Proposed Thresholding Function

We propose a custom thresholding function derived from the $\mu$-law and semisoft thresholding functions. Representing $\lambda(k)$ derived from (11) as $\lambda_1(k)$ and letting $\lambda_2(k) = 2\lambda_1(k)$, the proposed thresholding function is

developed as

$$(Y_{k,m})_{PCT} = \begin{cases} \alpha(k,m)sgn(Y_{k,m}) \times \Pi, \text{if } |(Y_{k,m})| < \lambda_1(k), \\ Y_{k,m}, \text{if } |(Y_{k,m})| > \lambda_2(k), \\ (1 - \alpha(k,m))\Psi_1 + \alpha(k,m)\Psi_2, \text{otherwise}, \end{cases} \tag{14}$$

where $\lambda_1(k)$ is $\lambda(k)$ derived from (11), $\lambda_2(k) = 2 \times \lambda_1(k)$ and

$$\Pi = (1 + \mu)^{\frac{|Y_{k,m}|}{\lambda_1(k)}} - 1, \tag{15}$$

$$\Psi_1 = sgn(Y_{k,m}) \times \lambda_2(k)\frac{|(Y_{k,m})| - \lambda_1(k)}{\lambda_2(k) - \lambda_1(k)}, \tag{16}$$

$$\Psi_2 = Y_{k,m}. \tag{17}$$

In (14), $(Y_{k,m})_{PCT}$ stands for the PWP coefficients thresholded by the proposed custom thresholding function and shape parameter of the proposed thresholding function is represented by $\alpha(k,m)$, which is taken as a constant in the proposed method.

The comparison of the proposed custom thresholding function with the conventional $\mu$-law and semisoft thresholding functions is shown in Fig.13. In the region between $\lambda_1$ and $\lambda_2$, this figure demonstrates the flexibility of the proposed thresholding operation in a sense that it can be viewed as $(1 - \alpha(k,m))(Y_{k,m})_{SS} + \alpha(k,m)(Y_{k,m})_{ML}$ which is a linear combination of the $\mu$-law and the semisoft thresholding function. Here, $(Y_{k,m})_{ML}$ stands for the PWP coefficients thresholded by the $\mu$-law thresholding function and $(Y_{k,m})_{SS}$ represents the PWP coefficients thresholded by the semisoft thresholding function. Unlike these functions, depending on the value of shape parameter $\alpha(k,m)$, it can be verified from (14) that the proposed thresholding function gets the following forms,

$$\lim_{\alpha(k,m)\to0} (Y_{k,m})_{PCT} = (Y_{k,m})_{SS},$$

$$\lim_{\alpha(k,m)\to1} (Y_{k,m})_{PCT} = (Y_{k,m})_{ML}.$$

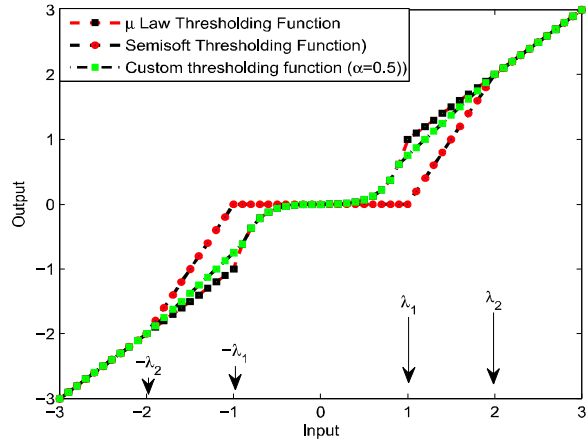

Fig. 13: Input Output Relation for semisoft, $\mu$-law and proposed custom thresholding function

*F. Inverse Perceptual Wavelet Packet Transform*

For a noisy speech frame, we obtain thresholded PWP coefficients using the proposed threshold in (11) and the proposed thresholding function in (14). An enhanced speech frame $\hat{r}[n]$ is synthesized by performing inverse PWP transform as

$\hat{r}[n] = PWP^{-1}(Y_{k,m})_{PCT}$.

The enhanced speech signal is reconstructed by using the standard overlap-and-add method [1].

TABLE I: Value of the shape parameter

| Shape parameter | Value |
|---|---|
| $\alpha$ | 0.5 |

## III. RESULTS

In this Section, a number of simulations is carried out to evaluate the performance of the proposed method.

### A. Simulation Conditions

Real speech sentences from the NOIZEUS database are employed for the experiments, where the speech data is sampled at 8 KHz [25]. To imitate a noisy environment, noise sequence is added to the clean speech samples at different SNR levels ranging from 15 dB to -15 dB. As in [26], two different types of noises, such as car and babble are adopted from the NOIZEUS databases [25].

In order to obtain overlapping analysis frames, hamming windowing operation is performed, where the size of each of the frame is 512 samples with 50% overlap between successive frames. A 6-level PWP decomposition tree with 10 db bases function is applied on the noisy speech frames [20], [27] resulting in subbands $k = 1, 2, .....24$. The value of the shape parameter used in the proposed method is given in Table I.

### B. Comparison Metrics

Standard Objective metrics namely, Segmental SNR (SNRSeg) improvement in dB, Perceptual Evaluation of Speech Quality (PESQ) and Weighted Spectral Slope (WSS) are used for the evaluation of the proposed method [2]. The proposed method is subjectively evaluated in terms of the spectrogram representations of the clean speech, noisy speech and enhanced speech. Formal listening tests are also carried out in order to find the analogy between the objective metrics and the subjective sound quality. The performance of our method is compared with some of the state-of-the-art speech enhancement methods, such as Universal [13] and SMPO [26] in both objective and subjective senses.

### C. Objective Evaluation

*1) Results for Speech signals with Car Noise:* SNRSeg improvement, PESQ and WSS for speech signals corrupted with car noise for Universal, SMPO and proposed method are shown in Fig.14, Table II and Fig.15.

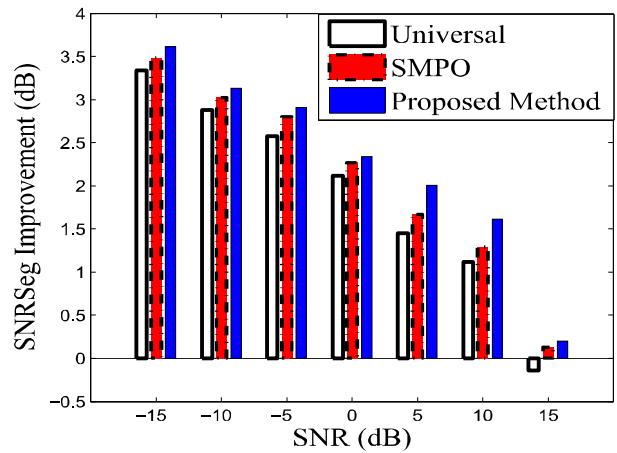

Fig. 14: SNRSeg Improvement for different methods in car noise

In Fig.14, the performance of the proposed method is compared with that of the other methods at different levels of SNR for car noise in terms of SNRSeg improvement. We see that the SNRSeg improvement increases as SNR decreases. At a low SNR of $-15dB$, the proposed method yields the highest SNRSeg improvement. Such larger values of SNRSeg improvement at a low level of SNR attest the capability of the proposed method in producing enhanced speech with better quality for speech severely corrupted by car noise.

TABLE II: PESQ for different methods in car noise

| SNR(dB) | Universal | SMPO | Proposed Method |
|---------|-----------|------|-----------------|
| -15 | 1.16 | 1.15 | 1.27 |
| -10 | 1.23 | 1.37 | 1.45 |
| -5 | 1.32 | 1.51 | 1.61 |
| 0 | 1.43 | 1.69 | 1.79 |
| 5 | 1.69 | 2.07 | 2.13 |
| 10 | 1.93 | 2.38 | 2.44 |
| 15 | 2.14 | 2.60 | 2.79 |

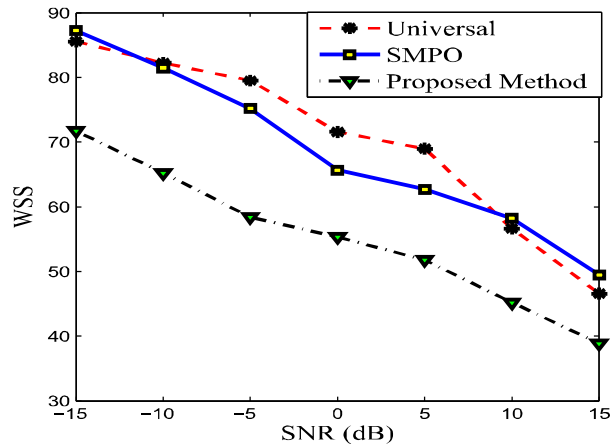

Fig. 15: WSS for different methods in car noise

In Table II, it can be seen that at a low level of SNR, such as $-15dB$ , all the methods show lower values of PESQ scores, whereas the PESQ score is much higher, as expected, for the proposed method. The proposed method also yields larger PESQ scores compared to that of the other methods at higher levels of SNR. Since, at a particular SNR, a higher PESQ score indicates a better speech quality, the proposed method is indeed better in performance in the presence of a car noise.

Fig.15 represents the WSS values as a function of SNR for the proposed method and that for the other methods. As shown in the figure, the WSS values resulting from all other methods are relatively larger for a wide range of SNR levels, whereas the proposed method is capable of producing enhanced speech with better quality as it gives lower values of WSS even at a low SNR of $-15dB$.

*2) Results for Speech signals with Multi-talker Babble Noise:* SNRSeg improvement, PESQ and WSS for speech signals corrupted with babble noise for Universal, SMPO and proposed methods are shown in Fig.16, 17 and 18, respectively.

In Fig. 16, it can be seen that at a low level of SNR of $-15dB$, the proposed method provides a SNRSeg improvement that is significantly higher than that of the methods of comparison. The proposed method still shows better performance in terms of SNRSeg improvement for higher SNRs also.

For speech corrupted with babble noise, in Fig.17, the mean values of PESQ with standard deviation obtained using the proposed method is plotted and compared with that of the other methods. From this plot, it is seen that over the whole SNR range considered, the proposed method continue to provide higher PESQ with almost non-overlapping standard deviation in the presence of babble noise.

The performance of the proposed method is compared with that of the other methods in terms of WSS in Fig.18 at different levels of SNRs in presence of babble noise. It is clearly seen from this figure that WSS increases as SNR decreases. At a low SNR of $-15dB$, the proposed method yields a WSS that is significantly lower than that of all other methods, which remains lower over the higher SNRs also.

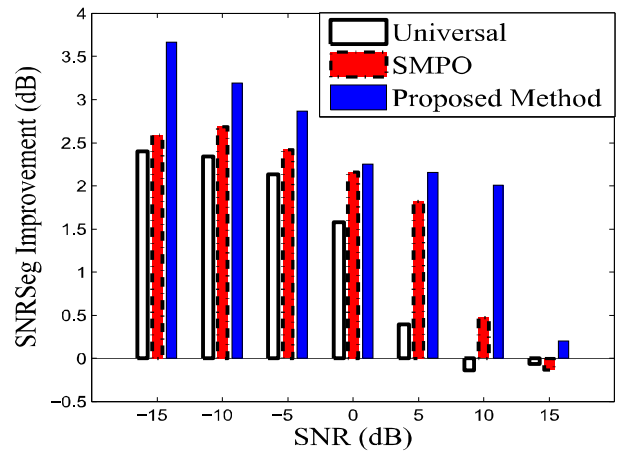

Fig. 16: SNRSeg Improvement for different methods in babble noise

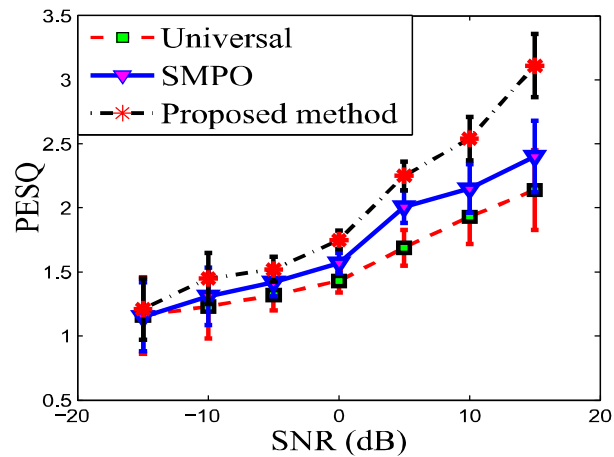

Fig. 17: PESQ for different methods in babble noise

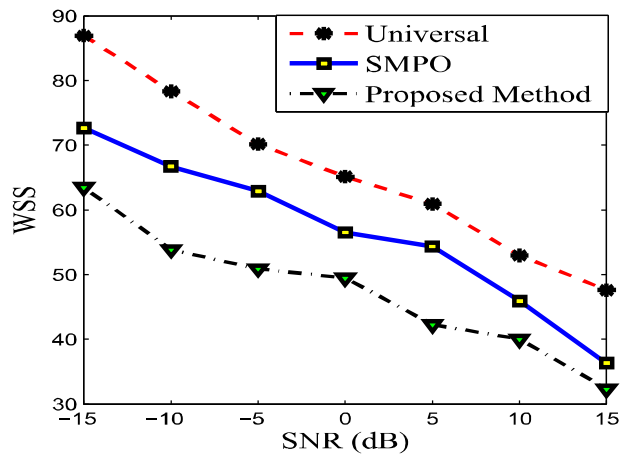

Fig. 18: WSS for different methods in babble noise

## D. Subjective Evaluation

In order to evaluate the subjective observation of the enhanced speech, spectrograms of the clean speech, the noisy speech, and the enhanced speech signals obtained by using the proposed method and all other methods are presented in Fig. 19 for car noise corrupted speech at an SNR of 10 dB. It is evident from this figure that the harmonics are well preserved and the amount of distortion is greatly reduced in the proposed method. Thus, the spectrogram observations with lower distortions also validate our claim of better speech quality as obtained in our objective evaluations in terms of higher SNR improvement in dB, higher PESQ score and lower WSS in comparison to the other methods. Another set of spectrograms for babble noise corrupted speech at an SNR of 10 dB is also presented in Fig.20. This figure attests that the proposed method has a better efficacy in preserving speech harmonics even in case of babble noise.

TABLE III: Mean Score of SIG scale for different methods in presence of car noise at 5 db

| Listener | Universal | SMPO | Proposed Method |
|---|---|---|---|
| 1 | 3.6 | 4.0 | 4.0 |
| 2 | 3.3 | 3.9 | 3.7 |
| 3 | 3.9 | 4.0 | 4.2 |
| 4 | 3.4 | 4.2 | 4.5 |
| 5 | 3.2 | 3.8 | 4.0 |
| 6 | 2.9 | 3.6 | 3.9 |
| 7 | 3.8 | 3.8 | 4.2 |
| 8 | 3.5 | 3.7 | 4.2 |
| 9 | 3.5 | 3.9 | 3.8 |
| 10 | 3.7 | 3.9 | 4.0 |

TABLE IV: Mean Score of BAK scale for different methods in presence of car noise at 5 db

| Listener | Universal | SMPO | Proposed Method |
|---|---|---|---|
| 1 | 4.0 | 4.5 | 5.0 |
| 2 | 4.3 | 4.9 | 4.7 |
| 3 | 4.2 | 4.4 | 4.9 |
| 4 | 4.4 | 4.7 | 4.8 |
| 5 | 4.2 | 4.8 | 4.7 |
| 6 | 3.9 | 4.6 | 4.9 |
| 7 | 3.8 | 3.9 | 4.4 |
| 8 | 4.4 | 4.6 | 4.6 |
| 9 | 3.5 | 3.8 | 4.5 |
| 10 | 4.2 | 4.5 | 4.8 |

Formal listening tests are also conducted, where ten listeners are allowed and arranged to perceptually evaluate the enhanced speech signals. A full set (thirty sentences) of the NOIZEUS corpus was processed by Universal, SMPO and proposed method for subjective evaluation at different SNRs. Subjective tests were performed according to ITU-T recommendation P.835 [25]. In this tests, a listener is instructed to successively attend and rate the enhanced speech signal based on (a) the speech signal alone using a scale of SIG (1 = very unnatural, 5 = very natural), (b) the background noise alone using a scale of background conspicuous/ intrusiveness (BAK) (1 = very conspicuous, very intrusive; 5 = not noticeable), and (c) the overall effect using the scale of the mean opinion score (OVRL) (1 = bad, 5 = excellent). More details about the testing methodology can be found in [28]. The mean scores of SIG, BAK, and OVRL scales for the three speech enhancement methods evaluated in the presence of car noise at an SNR of 5 dB are shown in Tables III, IV, and V. For the three methods evaluated using babble noise-corrupted speech at an SNR of 10 dB, the mean scores of SIG, BAK, and OVRL scales are also summarized in Tables VI, VII, and VIII. The mean scores in the presence of both car and babble noises demonstrate that the lower signal distortion (i.e., higher SIG scores) and the lower noise distortion (i.e., higher BAK scores) are obtained with the proposed method relative to that obtained by Universal and SMPO methods in most of the conditions. It is also shown that a consistently better performance in OVRL scale is offered by the proposed method not only in car but also in babble noisy conditions at both SNR levels considered in comparison to that provided by all the methods mentioned above. Overall, it is found that the proposed method possesses the highest subjective sound quality in comparison to that of the other methods in case of different noises at various levels of SNRs.

TABLE V: Mean Score of OVL scale for different methods in presence of car noise at 5 db

| Listener | Universal | SMPO | Proposed Method |
|---|---|---|---|
| 1 | 2.6 | 4.0 | 4.1 |
| 2 | 3.3 | 3.8 | 3.7 |
| 3 | 3.9 | 4.1 | 4.3 |
| 4 | 3.6 | 4.2 | 4.2 |
| 5 | 3.3 | 3.9 | 4.1 |
| 6 | 3.9 | 4.6 | 4.9 |
| 7 | 3.8 | 3.8 | 4.3 |
| 8 | 3.6 | 4.1 | 4.2 |
| 9 | 3.5 | 4.5 | 4.7 |
| 10 | 3.9 | 4.6 | 4.8 |

TABLE VI: Mean Score of SIG scale for different methods in presence of Babble noise at 5 db

| Listener | Universal | SMPO | Proposed Method |
|---|---|---|---|
| 1 | 3.6 | 4.0 | 4.0 |
| 2 | 3.3 | 3.9 | 3.7 |
| 3 | 3.9 | 4.0 | 4.2 |
| 4 | 3.4 | 4.2 | 4.5 |
| 5 | 3.2 | 3.8 | 4.0 |
| 6 | 2.9 | 3.6 | 3.9 |
| 7 | 3.8 | 3.8 | 4.2 |
| 8 | 3.4 | 3.6 | 4.1 |
| 9 | 3.5 | 3.9 | 3.7 |
| 10 | 3.7 | 3.8 | 3.9 |

TABLE VII: Mean Score of BAK scale for different methods in presence of Babble noise at 5 db

| Listener | Universal | SMPO | Proposed Method |
|---|---|---|---|
| 1 | 2.6 | 4.0 | 4.1 |
| 2 | 3.3 | 3.8 | 3.7 |
| 3 | 3.9 | 4.1 | 4.3 |
| 4 | 3.6 | 4.2 | 4.2 |
| 5 | 3.3 | 3.9 | 4.1 |
| 6 | 3.9 | 4.6 | 4.9 |
| 7 | 3.8 | 3.8 | 4.3 |
| 8 | 3.6 | 4.1 | 4.2 |
| 9 | 3.5 | 4.5 | 4.7 |
| 10 | 3.9 | 4.8 | 4.9 |

TABLE VIII: Mean Score of OVL scale for different methods in presence of Babble noise at 5 db

| Listener | Universal | SMPO | Proposed Method |
|---|---|---|---|
| 1 | 2.6 | 4.0 | 4.1 |
| 2 | 3.3 | 3.8 | 3.7 |
| 3 | 3.9 | 4.1 | 4.3 |
| 4 | 3.6 | 4.2 | 4.2 |
| 5 | 3.3 | 3.9 | 4.1 |
| 6 | 3.9 | 4.6 | 4.9 |
| 7 | 3.8 | 3.8 | 4.3 |
| 8 | 3.6 | 4.1 | 4.2 |
| 9 | 3.5 | 4.5 | 4.7 |
| 10 | 3.9 | 4.8 | 4.9 |

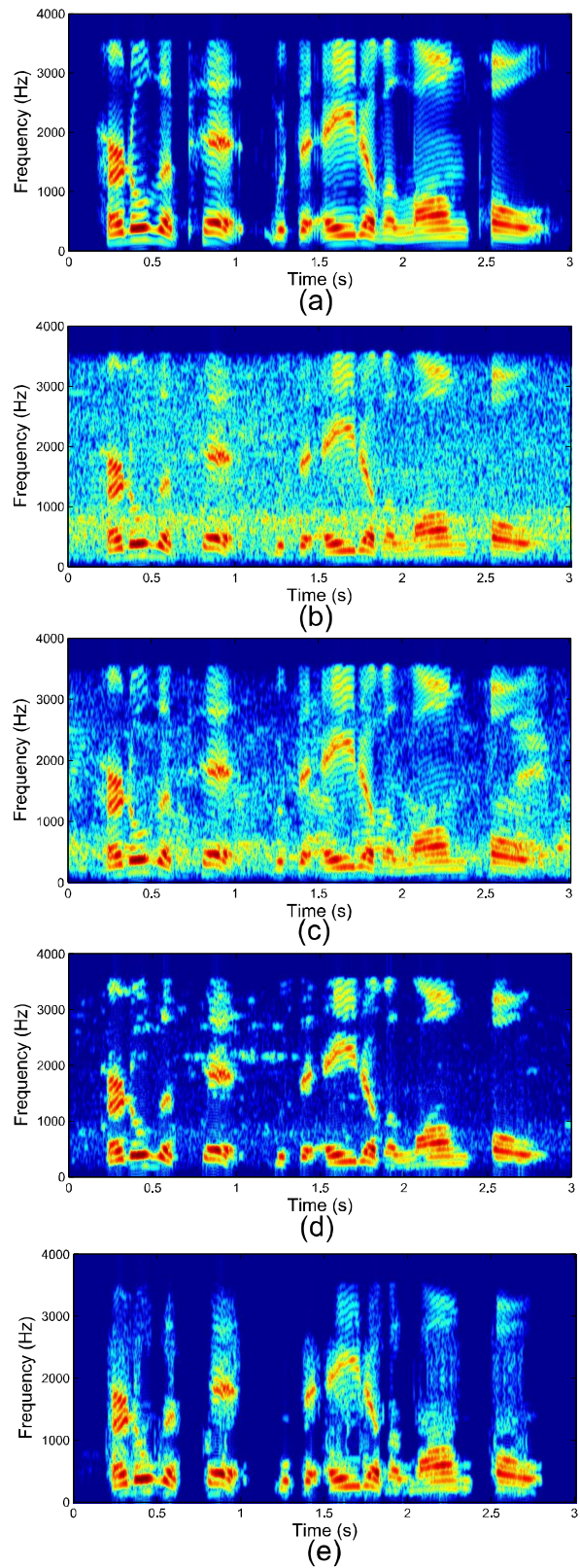

Fig. 19: Spectrograms of (a) Clean Signal (b) Noisy Signal with 10dB car noise; spectrograms of enhanced speech from (c) Universal method (d) SMPO method (e) Proposed Method

## IV. CONCLUSIONS

In order to obtain a suitable threshold value, we developed an exponential model-based technique employing the TE operated PWP coefficients of the noisy speech. The PWP coefficients of the noisy speech are thresholded by using the derived threshold in the proposed custom thresholding function. Extensive simulations and results show that the proposed method is capable of yielding consistently better results in terms of objective metrics, namely Segmental SNR Improvement in dB, PESQ, and WSS values in comparison to those of the existing methods. The clearly improved performance indicators of the proposed method are also shown in terms of much better spectrogram outputs and the higher scores in the formal subjective listening tests.

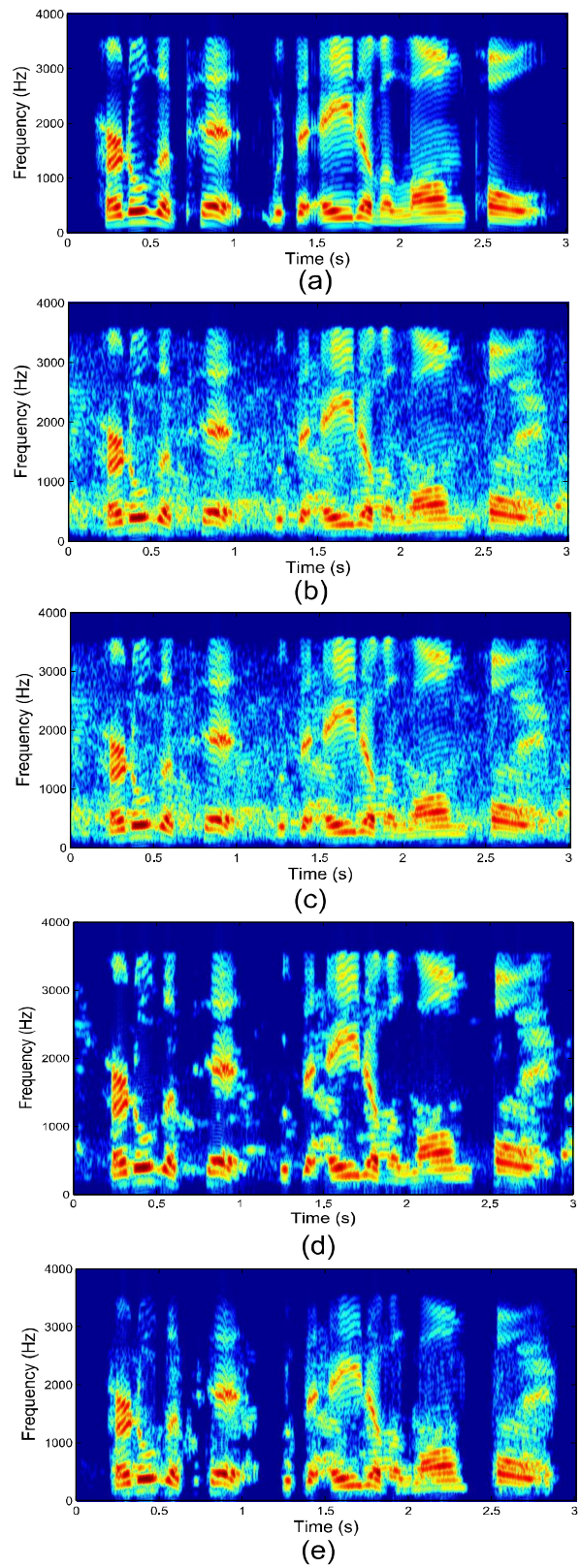

Fig. 20: Spectrograms of (a) Clean Signal (b) Noisy Signal with 10dB car noise; spectrograms of enhanced speech from (c) Universal method (d) SMPO method (e) Proposed Method